\documentclass[twocolumn,amssymb]{revtex4}
\usepackage{epsfig}
\usepackage{hyperref}
\usepackage{graphics}

\begin{document}
\title{Quantifying phases in homogenous twisted birefringent medium}
\author{{Dipti Banerjee$^*$ } and  {Srutarshi Banerjee $^*$}}
\email{$*$deepbancu@hotmail.com},\\
\email{$**$sruban.stephens@gmail.com}
\affiliation{$*$Department of Physics,Vidyasagar College for Women \\
39, Sankar Ghosh lane,Kolkata-700006, West Bengal,INDIA}\\
\affiliation{$**$Department of Electrical Engineering,Indian Institute of Technology \\Kharagpur-721302,West Bengal,INDIA}\\
\date{12.06.12}\\

\begin{abstract}
The internal birefringence of an optical medium develops the dynamical phase through natural rotation of incident polarized light. The uniform twist of the medium induces
 an external birefringence in the system.This can be visualized through the geometric phase by the solid angle in association with the angular twist per unit thickness of the medium $k$.An equivalent physical analysis in the $l=1$ orbital angular momentum sphere also
has been pointed out.
\end{abstract}

\maketitle

keywords: Birefringent,geometric phase.\\
Pacs code:$42.25.Bs$\\

The theory of twisted birefringent material was developed long back by Gennes \cite{paper1} and Chandrasekhar \cite{paper2} in connection with optics of cholesteric and twisted nematic liquid crystals.In birefringent media three kinds of twists are studied: (i) the limit of twist pitch is very short with respect to the optical wavelength \cite{paper3} (ii) the twist pitch is comparable to the optical wavelength that is solved by the previous approach. (iii) the very long twist pitch is known as Geometric Optics Approximation or the Maguin \cite{paper4} limit. Much later Jones \cite{paper5},\cite{paper6}, attracted our attention in formulating the transformation through optical elements such as linear and circular polarizer, retarder, rotator,etc arranged in proper sequence by the method of $(2\times2)$ matrix. Azzam studied later by differential matrix formalism \cite{paper7} the anisotropic behavior of the optical medium with and without depolarization.

The property of birefringence develops quantum phases in the optical material.
The appeared phases may be either dynamical or geometric or mixture of both.There are four types of Geometric Phases (GP) that have been reported in optics so far: i)The first identified GP is Pancharatnam phase \cite{paper8},$\Omega/2$, where $\Omega$ is the solid angle enclosed by the path on the Poincare sphere. Berry explained the quantal counterpart \cite{paper9} of Pancharatnam's phase in case of cyclic adiabatic evolution.
He also studied the phase two-form (GP) \cite{paper10} in connection with the dielectric tensor of the optical medium.
ii) The second kind of phase was experimentally performed by Chaio and co-workers \cite{paper11} when the light with fixed polarization slowly rotate around a closed path with varied directions. The developed GP was the spin-redirection or coiled light phase.
iii) The third one was developed by the squeezed state of light through the cyclic changes by Lorentz transformation \cite{paper12}.
iv) The fourth GP was studied by van Enk \cite{paper13}, in case of cyclic change of transverse mode pattern of Gaussian light beam without affecting the direction of propagation or the polarization of light.Bhandari \cite{paper14} studied the details of geometric phase in various combination of optical material.
Berry et.al. \cite{paper15} studied the phase two-form (GP) of the twisted stack of birefringent plates.

The physical mechanism of these different kind of geometric phases originate from spin or orbital angular momentum of polarized photon.The first observation of the angular momentum of light was performed by Beth \cite{paper16} through an experiment where a beam of right circularly polarized light was passed through a birefringent medium (quarter-wave plate) and transformed to left circularly polarized light.Indeed as pointed out by Enk \cite{paper13} that Pancharatnam phase in mode space is associated with spin angular momentum transfer of light and optical medium.The orbital angular momentum GP of polarized photon has been studied experimentally by Galvez et.al.\cite{paper17} in mode space and theoretically by Padgget \cite{paper18} in Poincare sphere.In recent days the OAM beams are generated by a kind of birefringent plate known as "q-plates", which are very fruitful application in classical and quantum regime.In an interesting method the physics of OAM beams by "q-plates" has been developed by Santamato et.al. \cite{paper19} recently.

All these recent findings indicate that our previous study on the GP of polarized photon
(passing through polarization matrix M and rotator) in connection with helicity, was an obvious new representation \cite{paper20}.Explicitly with the spinorial representation of polarized photon by spherical harmonics, we consider \cite{paper21} that as the light gets a fixed polarization, its helicity in connection with the spin angular momentum also fixed. It varies along with the rotation of plane of polarization of light.We have expanded this idea by finding the dielectric matrix and isolating GP in terms of helicity of polarized photon \cite{paper22}.In this paper the properties of the birefringent medium are visualized through the dynamical and geometric phases. Here it has been assumed that the polarized photon is passed through the uniformly twisted medium having very long twist pitch nearly at the Maguin limit.In the next section the Jones matrix representation of birefringent medium is reviewed while in section-2 the dynamical and geometric phase for slow uniform angle of twist has been evaluated.

\section{The matrix representation of birefringent medium}

An optical element could change the polarization of the incident polarized light. Jones
developed the effect by representation of $2\times2$ matrix \cite{paper5}
\begin{equation}
\vec{D}'=M\vec{D}
\end{equation}
If the input polarization of light is unaltered after going through an optical medium,
then the state can be identified as the eigenvector $\vec{D}_i$ of the
optical component $M_i$ following the eigen value equation
\begin{equation}
M_i\vec{D}_i=d_i\vec{D}_i
\end{equation}
where $d_i$ is the corresponding eigenvalue of a particular polarization matrix
 $M_i=\pmatrix{m_1&m_4\cr m_3&m_2}.$\\
The optical properties such as birefringence and dichroism of a homogeneous medium
can be described by the differential matrix $N$.Jones \cite{paper6} pointed out that $N$
 governs the propagation of the polarized light vector $\varepsilon$
 at a particular wavelength through an infinitesimal distance within the optical element
\begin{equation}
\frac{d\varepsilon}{dz}=\frac{dM}{dz}\varepsilon_0
=\frac{dM}{dz}M^{-1}\varepsilon=N\varepsilon
\end{equation}
where it is evident that $N$ is the operator that determines $dM/dz$ from polarization matrix $M$ as follows
\begin{equation}
N=\frac{dM}{dz}M^{-1}=\pmatrix{n_1 & n_2 \cr n_3 & n_4}
\end{equation}
If $N$ is independent of z, it shows its dependence on polarization matrix M \cite{paper6}
by
\begin{equation}
M=M_0\exp(\int{Ndz})
\end{equation}

In the lamellar representation suggested by Jones \cite{paper6}, a thin slab of a given medium is equivalent to a pile of retardation plates and partial polarizers. Eight constants are required to specify the real and imaginary part of the four matrix elements of N matrix ($2\times2$), each possessing one of the eight fundamental properties.The eight optical properties can be paired \cite{paper7} for following four properties.\\
i) Isotropic refraction and absorption\\
ii) Linear birefringence and linear dichroism along the xy coordinate axis.\\
iii) Linear birefringence and linear dichroism along the bisector of xy coordinate axes.\\
iv) Circular birefringence and circular dichroism.\\
The optical medium that has circular birefringence and linear birefringence will be our point of interest \cite{paper7}, and could have the following matrix form
\begin{eqnarray}
\theta_{cb}= \tau \pmatrix{0 & -1 \cr 1 & 0} \\
\theta_{lb}= \rho \pmatrix{0 & -i \cr i & 0}
\end{eqnarray}
These $\theta_{cb}$ and $\theta_{lb}$ matrices form the required differential matrix.
\begin{eqnarray}
N=\theta_{cb}+\theta_{lb}
=\pmatrix{0 & -\tau-i\rho \cr \tau+i\rho& 0}=\pmatrix{0 & n_2 \cr n_3 & 0}
\end{eqnarray}
where, $\tau$ is the circular birefringence that measures the rotation of the plane polarized light per unit thickness and $\rho$ is the part of linear birefringence that measures the difference between the two principal constants along the coordinate axes.

The evolution of the ray vector $\varepsilon={\varepsilon_1 \choose \varepsilon_2}$ as in eq.(3) passing through such medium $N$, could be re-written as
\begin{eqnarray}
\frac{d\varepsilon_1}{dz}=n_{1}\varepsilon_1 + n_{2}\varepsilon_2\\
\frac{d\varepsilon_2}{dz}=n_{3}\varepsilon_1 + n_{4}\varepsilon_2
\end{eqnarray}
For pure birefringent medium represented by eq.(8), one may use the evolution of ray vector
\begin{equation}
\frac{d\varepsilon_1}{dz}=n_2 \varepsilon_2,\\
\frac{d\varepsilon_2}{dz}=n_3 \varepsilon_1
\end{equation}
which implies that the spatial variation of component of electric vectors in one direction give the effect in the other perpendicular direction. Thus an exchange of optical power between the two component states of the polarized light takes place indicating the rotation of the ray vector after entering the medium.

Geometrically this state $\varepsilon$ is a point P on the surface of the Poincare sphere that defines a position vector $\vec{p}$ in three dimensional space. Huard pointed out  \cite{paper23} that the evolution of the vector $\vec{p}$ is equivalent to the cyclic change of the state vector during the passage of infinitesimal distance dz of the optical medium. The spatial change of vector as it passes through the crystal becomes
\begin{equation}
\frac{d\vec{p}}{dz}=\vec{\Omega} \times \vec{p}
\end{equation}
 A natural twist for an elementary angle $d\alpha=\Omega dz$ is experienced by the instantaneous vector $\vec{p}$ for thickness $dz$. The magnitude and direction of the rotation vector depends on thickness and inherent property of the optical medium.\\

Jones pointed out \cite{paper6} that as homogeneous birefringent crystal is uniformly twisted about the direction of transmission, the $N$ matrices are transformed upon rotation both for angle of twist dependent and independent of crystal thickness.In the former case, Jones mentioned the twisted matrix $N'$ in terms of untwisted matrix $N_0$ and rotation matrix.
\begin{equation}
N'=S(\theta)N_0 S(-\theta)
\end{equation}
  For angle of twist independent on crystal thickness,Jones showed that \cite{paper6} the twisted matrix $N'$ becomes
\begin{equation}
N^\prime=N_0 - kS(\pi/2)
\end{equation}
where $S(\pi/2)=\pmatrix{0 & -1 \cr 1 & 0 \cr}$ denotes the rotation matrix for normal incidence of light. The solution of the above eqs.(13) and (14) is $\varepsilon'=\exp(N'z)\varepsilon'_0$ where $\varepsilon'_0$ is the value of the vector $\varepsilon'$ at $z=0$.
We realize here from the definition of angle of twist per unit distance that k has similarity with $\Omega$ in eq.(12). The basic difference lies in their space of appearance where the former $(k)$ exist in the external and later $(\Omega)$ in the internal space.

In the next section we will point out that the nature of quantum phases will depend on the kind of angle of twist. Due to the property of inherent birefringence represented by $\eta$, a natural twist is realized by the incident polarized light acquiring dynamical phase. External twist of the optical medium develops further external birefringence of the medium by k that could be visualized by geometric phase(GP). It may be noted that GP will differ as and when the angle of twist is independent or dependent on the crystal thickness. From the work of Santamato \cite{paper19} one can realize that in former case GP is visualized in OAM sphere.

\section{Quantum phases by twisting homogeneous birefringent medium}

A polarized light traveling in the z direction can be written as a two component spinor
\begin{equation}
|\psi>={\psi_+ \choose \psi_-}
\end{equation}
in terms of the electric displacement vector $d_x,d_y$ where $\psi_{\pm}=(d_x \pm id_y)/\sqrt{2}$ .
Berry \cite{paper9} pointed out that the polarization matrix $M$ satisfying
$M|\psi>=1/2 |\psi>$, can be determined from the eigenvector $|\psi>$ using the relation $(|\psi><\psi|-1/2)$.

From the spherical harmonics, the eigenvector
\begin{equation}
|\psi> = {{Y_{1/2}}^{-1/2,1/2} \choose
{Y_{1/2}}^{-1/2,-1/2}} = {\cos\frac{\theta}{2}\exp i(\phi+\chi)/2
\choose \sin\frac{\theta}{2}\exp-i(\phi-\chi)/2}
\end{equation}
can be considered here omitting the phase factor $\exp-i(\phi-\chi)/2$, from above $|\psi>$ as
\begin{equation}
|\psi>={\cos\theta/2 e^{i\phi} \choose \sin\theta/2}
\end{equation}
In view of Berry \cite{paper9} the polarization matrix could be expressed as
\begin{equation}
\left.
\begin{array}{lcl}
M(r)&=& \frac{1}{2}{\psi_+ \choose \psi_-}(\psi_+~~~\psi_) - \frac{1}{2}\\
&=& \frac{1}{2}\pmatrix{{\psi_+\psi_+} & {\psi_+\psi_-} \cr {\psi_-\psi_+} & {\psi_-\psi_-}} -\frac{1}{2}\pmatrix{1 & 0 \cr 0 & 1}\\
&=& \frac{1}{2}\pmatrix{{\psi_+\psi_+ -1} & {\psi_+\psi_-} \cr {\psi_-\psi_+} & {\psi_-\psi_- -1}}
\end{array}
\right\}
\end{equation}
  Representing each term by spherical harmonics \cite{paper21} as
$${Y_1}^1 \approx \psi_- \psi_+ \approx {Y_{1/2}}^{-1/2,1/2}.{Y_{1/2}}^{-1/2,-1/2}$$
$${Y_1}^{-1} \approx \psi_+ \psi_- \approx {Y_{1/2}}^{1/2,1/2}.{Y_{1/2}}^{1/2,-1/2}$$
and
$${Y_1}^0 \approx (\psi_+ \psi_+ -1) \\or (1-\psi_- \psi_-) \\ \\
 \approx {Y_{1/2}}^{1/2,1/2}.{Y_{1/2}}^{-1/2,-1/2}$$
 $-{Y_{1/2}}^{-1/2,1/2}.{Y_{1/2}}^{1/2,-1/2}$\\
also $${Y_{1/2}}^{1/2,1/2}.{Y_{1/2}}^{-1/2,-1/2} \approx 1$$
 that results the polarization matrix in eq.(18) of the form
\begin{equation}
\left.
\begin{array}{lcl}
M(r)
&=& \frac{1}{2}\pmatrix{\cos\theta & \sin\theta e^{-i\phi}\cr
{\sin\theta e^{i\phi}}&{-\cos\theta}}
\end{array}
\right\}
\end{equation}
Every elements of the above polarization matrix eq.(19) can be realized as the product harmonics ${Y_1}^1$, ${Y_1}^{-1}$ and ${Y_1}^0$. This enable one to realize here the polarization matrix for orbital angular momentum $l=1$ \cite{paper21}.
\begin{equation}
M(r)=\frac{1}{2}\pmatrix{{Y_1}^0 & {Y_1}^1 \cr {Y_1}^-1 &{Y_1}^0\cr}\\
\end{equation}
The polarization matrix for this case $(l=1)$ parameterized by $(\theta,\phi)$ lies on the conventional Poincare sphere and equivalent as OAM sphere for $l=1$. Spin angular momentum (SAM) of polarized photon is associated with optical polarization. Another parameter $\chi$ for helicity is included that extend the Poincare sphere to $(\theta,\phi,\chi)$ whose picture is seen in fig.1. SAM space is possible to realize by parameters $(\theta,\chi)$.There are two eigenvalues of helicity
operator $+1$ and $-1$ that correspond to the right handed state (spin parallel to motion) and left handed state (spin opposite to motion) respectively. Hence the parameter for helicity, $\chi$ changes with the change of polarization of light
 For every OAM sphere there exist two SAM hemi-sphere. Since the eigen values of helicites for polarized photons is $\pm1$, the factor $1/2$ from the polarization matrix M has been omitted. For higher OAM states $l=2,3..$, further study is needed to evaluate the polarization matrix for a particular orbital angular momentum from the respective product harmonics ${Y_l}^m$.
\begin{figure}[ht]
\begin{center}
\resizebox{80mm}{!}{\includegraphics[scale=.4]{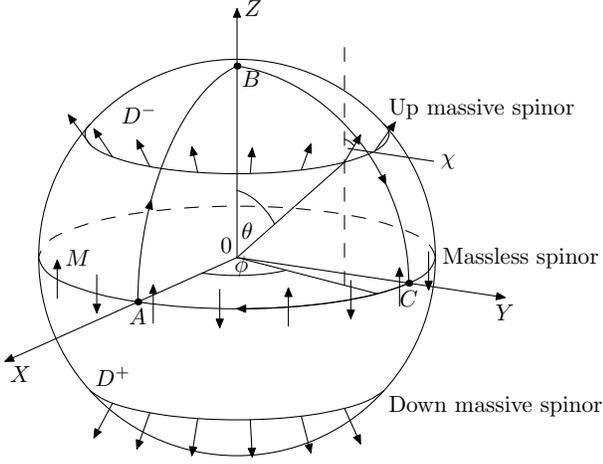}}
\end{center}
\caption{Spinorial representation of polarized photon}
\label{Fig:1}
\end{figure}
To study with helicity of photon the words of Berry \cite{paper10}, "Photons have no magnetic moment and so cannot be turned with a magnetic field but have the property of helicity to use" is very supportive. With this view we will study here further.

The property of birefringence of the optical medium can be represented by the differential matrix $N$. At a particular position $z$ of optical medium, the spatial variation of the polarization matrix $M$ becomes,
\begin{equation}
N=(\frac{dM}{d\theta})(\frac{d\theta}{dz})M^{-1}
\end{equation}
Considering $z=cos\theta$, the thickness of the optical medium, the N matrix can be obtained from M in eq.(19), where $\theta$ is the angular variable of light after refraction.
\begin{equation}
N=\eta \pmatrix{0 & -e^{i\phi} \cr e^{-i\phi} & 0 \cr}
\end{equation}
This N matrix has complex eigenvalues $\pm i\eta$. Comparing with our previous work \cite{paper21} it may be pointed out that $\eta$ has dependence on $\theta$ through the eigenvalue $1/sin\theta$ of the eigenvector,
\begin{equation}
\left(
\begin{array}{c}
 \pm i e^{i \phi } \\
 1
\end{array}
\right)
\end{equation}
Thus N matrix will be different for different values of $\theta$. The birefringence can not be measured if $\theta=0$ that makes $\eta=\infty$ where for $\theta=\pi/2$ and $\pi/3$ the accepted values of $\eta$ will be $1$ and $2$. Also its dependence on $\sin\theta$ indicate that $\eta$ might have only non-negative values. The nature of the optical medium could be identified comparing the above N matrix in eq.(22) with eq.(8). It is seen that developed N matrix is homogenous possessing both circular and linear birefringence represented by $\eta\cos\phi$ and $(-\eta\sin\phi)$ respectively.

When the angle of twist has dependence on the crystal thickness by $\theta=kz$,following Jones \cite{paper6} the twisted matrix $N'$ will be
\begin{equation}
{N}^\prime=\pmatrix {0 & -\eta e^{i\phi}+k \cr  \eta e^{-i\phi}-k & 0 \cr}
\end{equation}
The corresponding twisted ray will be obtained as the initial polarized light suffers rotation in opposite direction of the twisted matrix $N'$.
\begin{equation}
\varepsilon'=\pmatrix{\cos\theta & \sin\theta \cr -\sin\theta & \cos\theta \cr}
\left(
\begin{array}{c}
 i e^{i \phi }\\
 1
\end{array}
\right)
\end{equation}
in other words
\begin{equation}
\varepsilon'=
\left(
\begin{array}{c}
 ie^{i\phi}\cos\theta+\sin\theta \\
 -ie^{i\phi}\sin\theta+\cos\theta
\end{array}
\right)
\end{equation}.

Light having fixed polarization and helicity if suffers the slow variation of path in real space it can be mapped on to the surface of unit sphere in the wave vector space. Slow twist pitch as considered here is comparable with the Maguin limit \cite{paper4} in connection with the optics of cholesteric and twisted nematic liquid crystal.

The effect of twist help to achieve the geometric phase by the initial state $|A>$ of polarized light as it unite with the final $|A'>$.
\begin{equation}
<A|A'>=\pm \exp(i\gamma(C)/2)
\end{equation}
where $\gamma$ is the solid angle swept out on its unit sphere.

This work is based on the consideration of polarized light passing normally
through a medium having linear and circular birefringence. The incident polarized light suffers a natural twist due to inherent property of the medium. As a result the dynamical phase $\gamma$ is developed in the optical medium $N$. Further external twist with the consideration of $d\theta=d\theta'$, develops the phase $\gamma'$ that has dynamical and geometrical part too.
\begin{eqnarray}
\gamma=\varepsilon^* \frac{d\varepsilon}{d\theta}\frac{d\theta}{dz}={\varepsilon^*}N\varepsilon\\
\gamma'={\varepsilon'}^* \frac{d\varepsilon'}{d\theta}\frac{d\theta}{dz}={\varepsilon'}^* N'{\varepsilon'}
\end{eqnarray}
The dynamical phase $\gamma$ can be obtained using eq.(22) and (23) in (28)
\begin{eqnarray}
\gamma={\varepsilon^*}N\varepsilon
&=&{\varepsilon_1}^*n_2\varepsilon_2+{\varepsilon_2}^*n_3\varepsilon_1\nonumber\\
&=&(-ie^{-i\phi})(-\eta e^{i\phi})+(\eta e^{-i\phi})(ie^{i\phi})\nonumber\\
&=&2i\eta
\end{eqnarray}
Comparing this eq.(30) with (27), the developed dynamical phase $\gamma$ appears as imaginary term in exponent. It could be varied between $1 \& 2$ for positive $\theta$ values through $1/\sin\theta$. Hence Fig.-2 shows the uniform variation of this dynamical phase $\gamma$ with the natural birefringence of the medium $\eta$. If $\theta$ becomes negative, the corresponding $\eta$ shows negative value also.
\begin{figure}[h]
\begin{center}
\resizebox{60mm}{!}{\includegraphics[scale=.4]{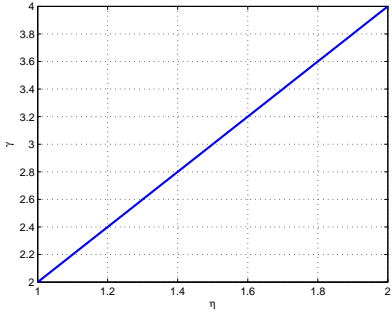}}
\end{center}
\caption{Variation of dynamical phase $\gamma$}
\label{Fig:2}
\end{figure}
Any ray passing through $N$ or $N'$ will suffer a natural twist due to the internal dynamics of the birefringent medium. The polarized light passing through the twisted medium $N'$ will acquire the phase $\gamma'$, that has two parts, one from the dynamics and another through the parametric change of the medium. The phase $\gamma'$ will contain both the dynamical and geometric phase in the exponent. To grasp the geometric phase due to external twist one could make the difference $\gamma'-\gamma$ to eliminate the dynamical phase developed.
\begin{figure}[h]
\begin{center}
\resizebox{60mm}{!}{\includegraphics[scale=.4]{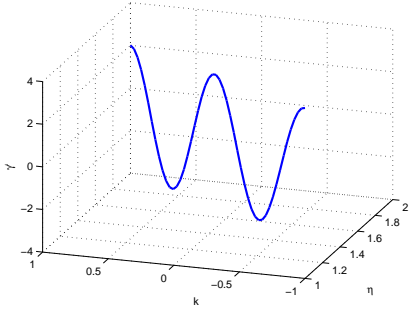}}
\end{center}
\caption{Variation of Net Phase $\gamma'$ with $\eta$ and k}
\label{Fig:3}
\end{figure}
To calculate the net quantum phase $\gamma'$ appeared after a twist, the twisted matrix $N'$ in eq.(24) is used which intuitively will act on the twisted light ray $\varepsilon'$ in eq.(26).
\begin{figure}[h]
\begin{center}
\resizebox{60mm}{!}{\includegraphics[scale=.4]{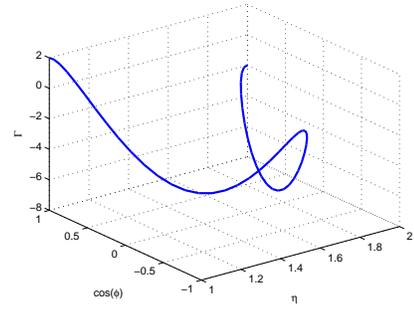}}
\end{center}
\caption{Variation of $\gamma'$ with k and $\cos(\phi)$.}
\label{Fig:4}
\end{figure}
\begin{eqnarray}
&&\gamma'={{\varepsilon'}_1}^*{n'}_2{\varepsilon'}_2+{\varepsilon_2}^* {n'}_3 {\varepsilon'.}_1\nonumber\\
 &=&(-ie^{-i\phi}\cos\theta+\sin\theta)(k-\eta e^{i\phi})(-ie^{i\phi}\sin\theta+\cos\theta)\nonumber \\
&+&(ie^{-i\phi}\sin\theta+\cos\theta)(\eta e^{-i\phi}-k)(ie^{i\phi}\cos\theta+\sin\theta)\nonumber\\
&=&i[2\eta {\sin}^2 \theta \cos2\phi-2k\cos\phi]
\end{eqnarray}
The above phase $\gamma'$ at $\theta=\pi/2$ becomes
\begin{equation}
\gamma'=2i[\eta \cos2\phi-k\cos\phi]
\end{equation}
Here the respective variation of net phase $\gamma'$ with $\eta$ and k are shown in fig.3 and in fig.4.
\begin{figure}[h]
\begin{center}
\resizebox{60mm}{!}{\includegraphics[scale=.4]{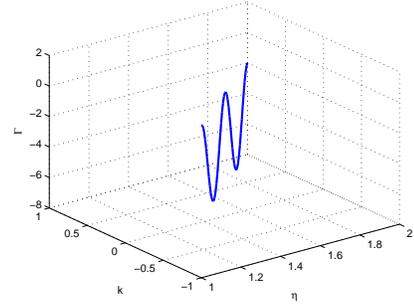}}
\end{center}
\caption{Variation of $\Gamma$ with $\eta$ and k.}
\label{Fig:5}
\end{figure}
As the dynamical phase is independent of twist angle $\theta$,the geometric phase in case of $\theta=\pi/2$ could be recovered by
\begin{equation}
\Gamma=\gamma'-\gamma
=i[2\eta(\sin^2 \theta \cos 2\phi-1)-2k\cos\phi]
\end{equation}
\begin{figure}[h]
\begin{center}
\resizebox{60mm}{!}{\includegraphics[scale=.4]{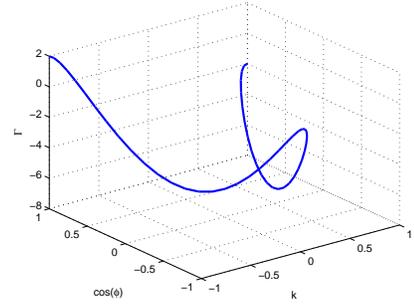}}
\end{center}
\caption{Variation of $\Gamma$ with $\cos \phi$ and k.}
\label{Fig:6}
\end{figure}
in other words
\begin{equation}
\Gamma=i\eta[(1-\cos 2\theta)\cos 2\phi-2]-2ik\cos\phi
\end{equation}
Similar nature of curve is seen for the net phase $\gamma'$ and geometric phase $\Gamma$. Fig.5 and Fig.6 shows the variation of $\Gamma$ with $\eta$ and k  and $\Gamma$ with $\cos\phi$ and k respectively.
\begin{figure}[h]
\begin{center}
\resizebox{60mm}{!}{\includegraphics[scale=.4]{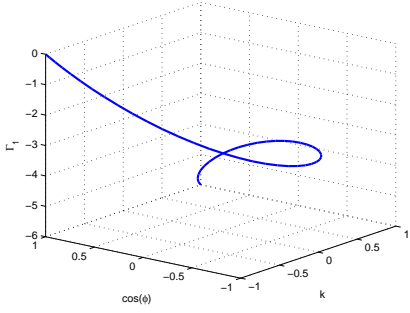}}
\end{center}
\caption{Variation of $\Gamma_1$ with $\cos \phi$ and k.}
\label{Fig:7}
\end{figure}
It may be noted that $\eta$ and $k$ parameters have dependence on the angle of incidence $\theta$ and angle of external twist $\theta'$ respectively. Here for simplicity the two angles ($\theta=\theta'$) are considered equal.
The angle $\phi$ is associated with the natural(internal) twist of the light ray inside the medium. The three types of phases $\gamma,\gamma'$ and $\Gamma$ could be studied graphically with respect to variation of k and $\eta$.
At normal incidence $\theta=\pi/2$,the GP will be
\begin{equation}
\Gamma=i[2\eta(\cos 2\phi-1)-2k\cos\phi]
\end{equation}
At $\theta=0$ angle of twist,GP has been identified as $\Gamma_1$
\begin{equation}
\Gamma_1=
-2i[\eta+k\cos\phi]
\end{equation}\\
Fig.7 and 8 shows the variation of GP at $\theta=0$ with respective parameters $\cos \phi$, $\eta$ and k.
\begin{figure}[h]
\begin{center}
\resizebox{60mm}{!}{\includegraphics[scale=.4]{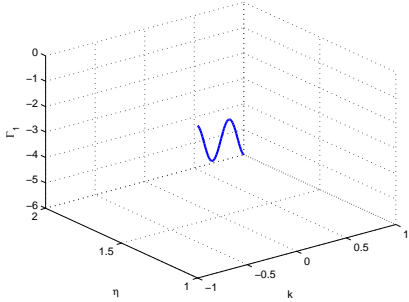}}
\end{center}
\caption{Variation of $\Gamma_1$ with $\eta$ and k.}
\label{Fig:8}
\end{figure}
The property of circular birefringence of the medium visualized by the parameter $\eta$ makes a natural twist to the incident light and for appearing dynamical phase. The external twist of the optical medium though is associated with $k$, the developed geometric phase has its dependence on $\eta$ also. Graphical analysis shows that the presence of external birefringence introduces a spiral behavior of the geometric phase. To realize GP in the OAM and SAM space the respective parameters $\phi$ and $\chi$ are responsible.

We like to extend our study using circularly polarized light incident on the birefringent medium.Let us identify the LCP and RCP of polarization of light by
\begin{equation}
|L>={1 \choose i},~~~~~|R>={1 \choose -i}
\end{equation}
The $|L>$ and $|R>$ states gives rise to the dynamical phases for up polarized photon
\begin{eqnarray}
\gamma_L = {<L|N|L>}=-2i\eta \cos\phi\\
\gamma_R = {<R|N|R>}=2i\eta \cos\phi
\end{eqnarray}
In case of twisted birefringent crystal, incidence of left circularly polarized light (LCP) on $N'$ develop the phase $\gamma'$
\begin{equation}
\begin{array}{lcl}
{\gamma'}_L = <L|N'|L>\\
=(1 -i)e^{-i\theta}\pmatrix {0 & -\eta e^{i\phi}+k \cr \eta e^{-i\phi}-k & 0 \cr}
{1  \choose  i}e^{i\theta}\\
= 2ik-2i\eta \cos\phi
\end{array}
\end{equation}
that consists both the dynamical and geometrical phases. The later(GP) could be isolated by
$\Gamma=\gamma'-\gamma$
\begin{equation}
\Gamma=2ik-2i\eta \cos\phi+2i\eta \cos\phi=2ik
\end{equation}
Here it seen that for circularly polarized light the dynamical phase and geometric phase depend only on the external birefringence k and the internal birefringence $\eta$ respectively. \\

\pagebreak
{\bf Discussion:}\\

In this communication two types of birefringence internal and external are studied.
Due to the property of inherent birefringence represented by $\eta$, a natural twist is realized by the incident polarized light acquiring dynamical phase. The dynamical phase $\gamma$ in all cases varies linearly with internal birefringence $\eta$ of the medium.
External twist of the optical medium develops further external birefringence of the medium by k that could be visualized by geometric phase(GP). It may be noted that GP will differ as and when the angle of twist is independent or dependent on the crystal thickness.
 GP has dependence on both the internal and external birefringence when eigen polarized light is passed through the twisted optical medium. It depends completely on the external birefringence k of the optical medium for the passage of left or right circularly polarized light. Further could be noted that if $\phi=0$ the value of geometric phase for the two types of polarized light becomes identical. In future we wish to study the twisted optical medium having the property of dichroism.\\ \\

{\bf Acknowledgement}: This work had been supported by the Abdus Salam International Centre for Theoretical Physics (ICTP), Trietse, Italy. Correspondence with Prof.Santamato,Napoli,Italy is gratefully acknowledged. Also the help from Mr.S.Bhar, colleague of DB at VCFW (Department of Physics) is acknowledged.

\pagebreak

\end{document}